\documentclass{article}

\usepackage[pdftex]{graphicx}
\usepackage[pdftex]{xcolor}
\usepackage[pdftex, hidelinks]{hyperref}

\usepackage{cite, url}

\usepackage{xparse, comment}

\usepackage[caption=false,font=footnotesize]{subfig}
\usepackage{array, amsmath, amssymb, amsfonts, bm, mathtools}
\usepackage{booktabs}
\usepackage{algorithm, algpseudocode}

\usepackage{spconf}

\newcommand{\norm}[1]{| #1 |}
\NewDocumentCommand\p{o m o}{p\hspace{-0.8pt}\IfValueT{#1}{_{#1}}\left( {#2} \IfValueT{#3}{\hspace{-0.8pt}\left| {#3} \right.} \hspace{-0.8pt}\right)}
\NewDocumentCommand\q{o m o}{q\hspace{-0.8pt}\IfValueT{#1}{_{#1}}\left( {#2} \IfValueT{#3}{\left| {#3} \right.} \right)}
\NewDocumentCommand\E{o m}{\mathbb{E}\IfValueT{#1}{_{#1}} \left[ #2 \right]}

\newcommand{\KL}[2]{\mathbb{KL}\left[ #1 \left| {#2} \right.\right]}

\newcommand{\distnormal}[2]{\mathcal{N}\hspace{-0.8pt}\left({#1}, {#2}\right)}
\newcommand{\distcmpnormal}[2]{\mathcal{N}_{\mathbb{C}}\hspace{-0.8pt}\left({#1}, {#2}\right)}
\newcommand{\distgamma}[2]{\mathcal{G}\hspace{-0.8pt}\left({#1}, {#2}\right)}
\newcommand{\distexp}[1]{\mathrm{Exp}\hspace{-0.8pt}\left({#1}\right)}
\newcommand{\distgig}[3]{\mathcal{GIG}\hspace{-0.8pt}\left({#1}, {#2}, {#3} \right)}

\NewDocumentCommand\newletter{m m o m m}{
\NewDocumentCommand#1{s t@ o}{%
\IfBooleanTF{##1}{\mathbf{\MakeUppercase{#2}}\IfValueT{#3}{^{#3}}}{%
\IfBooleanTF{##2}{{\bm #2}\IfValueT{#3}{^{#3}}_{\IfValueTF{##3}{##3}{#5}}}{%
{#2}\IfValueT{#3}{^{#3}}_{\IfValueTF{##3}{##3}{#4}}%
}}}}

\newletter{\x}{x}{ft}{t}
\newletter{\s}{s}{ft}{t}
\newletter{\n}{n}{ft}{t}

\newletter{\w}{w}{fk}{k}
\newletter{\h}{h}{kt}{k}
\newletter{\z}{z}{dt}{t}
\newletter{\zold}{z}{dt}{t}
\newletter{\zp}{z}[*]{dt}{t}

\newcommand{\auxphi}[1][ftk]{\phi_{#1}}

\newcommand{\auxlmd}[1][ft]{\lambda_{#1}}
 
\NewDocumentCommand\funmu{O{n} D(){\s[(1:F)t]}}{\mu_{#1}\left( #2 \right)}
\NewDocumentCommand\funsig{O{f} O{\z}}{\sigma_{#1}\left( #2 \right)}

\newcommand{\setR}{\mathbb{R}}
\newcommand{\setRp}{\mathbb{R}_+}
\newcommand{\setC}{\mathbb{C}}

\newcommand{\eye}{{\bf I}}

\newcommand{\argmax}{\operatornamewithlimits{argmax}}

\long\def\symbolfootnote[#1]#2{\begingroup\def\thefootnote{\fnsymbol{footnote}}
\footnote[#1]{#2}\endgroup}

\title{Statistical Speech Enhancement Based on Probabilistic Integration \\ of Variational Autoencoder and Non-Negative Matrix Factorization}

\name{
\shortstack{Yoshiaki Bando$^1$, Masato Mimura$^1$, Katsutoshi Itoyama$^1$, Kazuyoshi Yoshii$^{1,2}$, Tatsuya Kawahara$^1$}
\thanks{Thanks to JSPS KAKENHI No. 15J08765 for funding.}}
\address{%
  $^1$Graduate School of Informatics, Kyoto University, Sakyo-ku, Kyoto 606-8501, Japan \\%
  $^2$Center for Advanced Intelligence Project, RIKEN, Chuo-ku, Tokyo 103-0027, Japan}

\begin{document}
\ninept

\setlength\abovedisplayskip{1.1mm}
\setlength\belowdisplayskip{1.1mm}

\maketitle
\begin{abstract}
This paper presents a statistical method of single-channel speech enhancement that uses a variational autoencoder (VAE) as a prior distribution on clean speech.
A standard approach to speech enhancement is to train a deep neural network (DNN) to take noisy speech as input and output clean speech.
Although this supervised approach requires a very large amount of pair data for training, it is not robust against unknown environments.
Another approach is to use non-negative matrix factorization (NMF) based on basis spectra trained on clean speech in advance and those adapted to noise on the fly.
This semi-supervised approach, however, causes considerable signal distortion in enhanced speech due to the unrealistic assumption that speech spectrograms are linear combinations of the basis spectra.
Replacing the poor linear generative model of clean speech in NMF with a VAE---a powerful nonlinear deep generative model---trained on clean speech, we formulate a unified probabilistic generative model of noisy speech.
Given noisy speech as observed data, we can sample clean speech from its posterior distribution.
The proposed method outperformed the conventional DNN-based method in unseen noisy environments.
\end{abstract}

\begin{keywords}
  Single-channel speech enhancement, variational autoencoder, Bayesian signal processing
\end{keywords}


\section{Introduction}

Deep neural networks (DNNs) have demonstrated excellent performance in single-channel speech enhancement~\cite{lu2013speech, heymann2016neural, nugraha2016multichannel, narayanan2013ideal, pascual2017segan, wang2017recurrent}.
The denoising autoencoder (DAE)~\cite{lu2013speech}, for example, is a typical variant of such networks,
  which is trained to directly convert a noisy speech spectrogram to a clean speech spectrogram with a supervised training.
Alternatively, a DNN can be trained to predict time-frequency (TF) masks called ideal ratio masks (IRMs) that represent ratios of speech to input signals and are used for obtaining a speech spectrogram from a noisy spectrogram~\cite{narayanan2013ideal}.
Although it is necessary to prepare as training data a large amount of pairs of clean speech signals and their noisy versions,
  these supervised methods often deteriorate in unknown noisy environments.
This calls for semi-supervised methods that are trained by using only clean speech data in advance and then adapt to unseen noisy environments.

Statistical source separation methods based on the additivity of speech and noise spectrograms have also been used for speech enhancement~\cite{ephraim1984speech, loizou2013speech, mohammed2017statistical, smaragdis2007supervised, sun2015lowranksparsenmf}.
Non-negative matrix factorization (NMF)~\cite{fevotte2009nonnegative, mohammed2017statistical}, for example, regards a noisy speech spectrogram as a non-negative matrix and approximates it as the product of two non-negative matrices (a set of basis spectra and a set of the corresponding activations).
If a partial set of basis spectra is trained in advance from clean speech spectrograms,
  the noisy spectrogram is decomposed into the sum of speech and noise spectrograms in a semi-supervised manner.
Robust principal component analysis (RPCA)~\cite{sun2014noise, huang2012singing} is another promising method that can decompose a noisy spectrogram into a sparse speech spectrogram and a low-rank noise spectrogram in an unsupervised manner.
These conventional statistical methods, however, have a common problem that the linear representation or the sparseness assumption of speech spectrograms is not satisfied in reality and results in considerable signal distortion.

\begin{figure}[t]
  \centering
  \includegraphics[width=0.85\hsize]{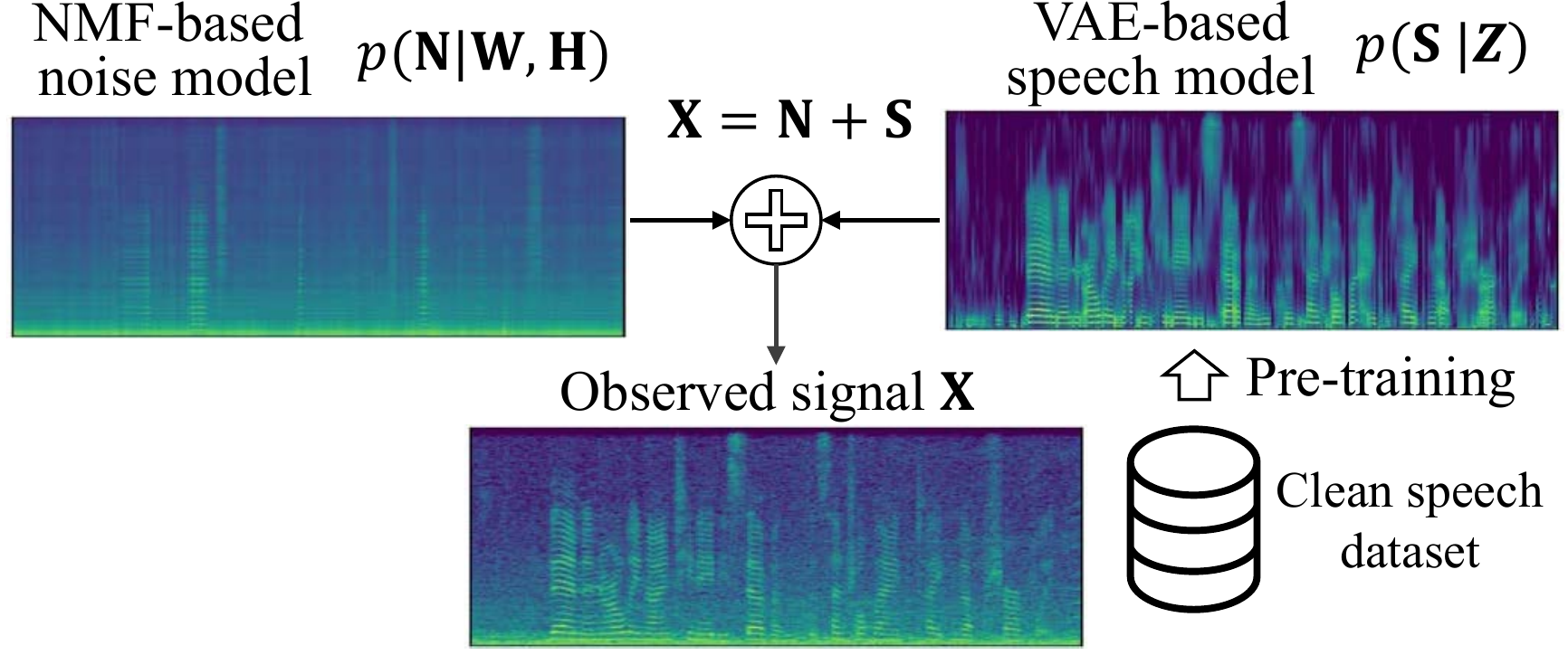}
  \vspace{-4mm}
  \caption{Overview of our speech enhancement model.}
  \label{fig:overview}
  \vspace{-4mm}
\end{figure}

\symbolfootnote[0]{\fontsize{7.5pt}{0pt}\selectfont \hspace{-9pt}*Demo page: \url{http://sap.ist.i.kyoto-u.ac.jp/members/yoshiaki/demo/vae-nmf/}}\hspace{-3pt}
Recently, deep generative models such as generative adversarial networks (GANs) and variational autoencoders (VAEs) have gained a lot of attention for learning a probability distribution over complex data (e.g., images and audio signals) that cannot be represented by conventional linear models~\cite{goodfellow2014generative, kingma2013auto, fabius2014variational, hsu2017learning, blaauw2016modeling}.
GANs and VAEs are both based on two kinds of DNNs having different roles.
In GANs~\cite{goodfellow2014generative}, a {\it generator} is trained to synthesize data that fool a {\it discriminator} from a latent space while the discriminator is trained to detect synthesized data in a minimax-game fashion.
In VAEs~\cite{kingma2013auto, fabius2014variational}, on the other hand, an {\it encoder} that embeds observed data into a latent space and a {\it decoder} that generates data from the latent space are trained jointly such that the lower bound of the log marginal likelihood for the observed data is maximized. 
Although in general GANs can generate more realistic data,
  VAEs provide a principled scheme of inferring the latent representations of both given and new data.

In this paper we propose a unified probabilistic generative model of noisy speech spectra by combining a VAE-based generative model of speech spectra with an NMF-based generative model of noise spectra (Fig.~\ref{fig:overview}).
The VAE is trained in advance from a sufficient amount of clean speech spectra and its decoder is used as a prior distribution on clean speech spectra included in noisy speech spectra.
Given observed data, we can estimate both the latent representations of speech spectra as well as the basis spectra and their activations of noise spectra through Bayesian inference based on a Markov chain Monte Carlo (MCMC) initialized by the encoder of the VAE. Our Bayesian approach can adapt to both unseen speech and noise spectra by using prior knowledge of clean speech and the low-rankness assumption on noise instead of fixing all the parameters in advance.


\section{Related work}
\vspace{-0.5mm}
This section overviews DNN-based speech enhancement and introduces the variational autoencoder (VAE).

\vspace{-0.5mm}
\subsection{DNN-based speech enhancement}
\vspace{-0.5mm}
Various network architectures and cost functions for enhancing speech signals have been reported~\cite{lu2013speech, heymann2016neural, nugraha2016multichannel, narayanan2013ideal, pascual2017segan, wang2017recurrent}.
The popular approach of DNN-based speech enhancement is to train a DNN to directly represent clean speech~\cite{wang2017recurrent}.
The DNN is trained using simulated noisy data constructed by adding noise to speech as input and clean speech as the target.
There are several methods that combine a supervised NMF and a DNN~\cite{kang2015nmf, vu2016combining}.
A DNN is trained to estimate activation vectors of the pre-trained basis vectors corresponding to speech and noise.
Bayesian WaveNet~\cite{qian2017speech} uses two networks: one, called a prior network, represents how likely a signal is speech and the other, called a likelihood network, represents how likely a signal is included in the observation.
These two networks enhance the noisy speech signal with a maximum a posteriori (MAP) estimation.
Another reported method uses two networks that are trained to represent how likely the input signal is speech or noise, respectively~\cite{grais2014deep}.
The speech signal is enhanced by optimizing a cost function so that the estimated speech maximizes the speech-likelihood network and minimizes the noise-likelihood network.
All the above mentioned methods are trained with datasets of both speech and noise signals.
A DNN-based method using only training data of speech signals was reported~\cite{sun2016unseen}.
This method represents speech and noise spectra with two autoencoders (AEs).
The AE for speech is pre-trained, whereas that for noise is trained at the inference for adapting to the observed noise signal.
Since the inference of this framework is under-determined,
  the estimated speech is constrained to be represented by a pre-trained NMF model.
It, thus, might have the same problem as the semi-supervised NMF.

\vspace{-0.5mm}
\subsection{Variational autoencoder}\label{sec:vae}
\vspace{-0.5mm}
A VAE~\cite{kingma2013auto} is a framework for learning the probability distribution of a dataset.
In this subsection, we denote by $\x*$  a dataset that contains $F$-dimensional samples $\x@[t] \in \setR^{F}$ ($t = 1, \ldots, T$).
The VAE assumes that a $D$-dimensional latent variable (denoted by $\z@[t] \in \setR^D$)\\ follows a standard Gaussian distribution and each sample $\x@[t]$ is stochastically generated from a conditional distribution $\p{\x@[t]}[\z@[t]]$:
\begin{align}
  \z@[t] &\sim \distnormal{\bf 0}{{\bf I}_D}, \\
  \x@[t] &\sim \p{\x@[t]}[\z@[t]],
\end{align}
where $\distnormal{\mu}{\sigma}$ represents a  Gaussian distribution with mean parameter $\mu$ and variance parameter $\sigma$.
$\p{\x@[t]}[\z@[t]]$ is called a decoder and parameterized as a well-known probability density function whose parameters are given by nonlinear functions represented as neural networks.
For example, Kingma et al.~\cite{kingma2013auto} reported a VAE model that has the following Gaussian likelihood function:
\begin{align}
  \hspace{-1mm}\x@[t] \sim \p{\x@[t]}[\z@[t]] = \prod_f \p{\x[ft]}[\z@[t]] = \prod_f \distnormal{\mu^x_{f}(\z@[t])}{\sigma^x_{f}(\z@[t])},
\end{align}
where $\mu^x_{f}: \setR^D\rightarrow \setR$ and $\sigma^x_{f}: \setR^D\rightarrow \setRp$ are neural networks representing the mean and variance parameters, respectively.

The objective of VAE training is to find a likelihood function $\p{\x@[t]}[\z@[t]]$ that maximizes the log marginal likelihood:
\begin{align}
  \argmax_{\p{\x@[t]}[\z@[t]]} \log \p{\x*} = \argmax_{\p{\x@[t]}[\z@[t]]} \prod_t \int \p{\x@[t]}[\z@[t]] \p{\z@[t]} d\z@[t].
\end{align}
Since calculating this marginal likelihood is intractable, it is approximated with a variational Bayesian (VB) framework.
The VAE first approximates the posterior distribution of $\z@[t]$ with the following variational posterior distribution $\q{\z@[t]}$ called an encoder:
{\setlength\belowdisplayskip{0.3mm}
\begin{align}
  \p{\z@[1], \ldots, \z@[T]}[\x*] &\approx \prod_t \q{\z@} = \prod_{d,t} \q{\z} \\
  &= \prod_{d,t} \distnormal{\mu^z_d(\x@)}{\sigma^z_d(\x@)},
\end{align}}%
where $\mu^z_d: \setR^F\rightarrow\setR$ and $\sigma^z_d: \setR^F\rightarrow\setRp$ are nonlinear functions representing the mean and variance parameters, respectively.
These functions are formulated with DNNs.
By using the variational posterior,
  the log marginal likelihood is lower-bounded as follows:
{\setlength\belowdisplayskip{0.3mm}
\begin{align}
  \hspace{-2mm} \log \p{\x*} &= \sum_t \log \int \p{\x@}[\z@]\p{\z@} d\z@ \\
  &\geq \sum_t \int \q{\z@} \log \frac{\p{\x@}[\z@]\p{\z@}}{\q{\z@}} d\z@ \\
  &= -\sum_t \KL{\q{\z@}}{\p{\z@}} + \sum_k \E[q]{\log \p{\x@}[\z@]}, \label{eq:vae-elbo}
\end{align}}%
where $\KL{\cdot}{\cdot}$ represents the Kullback-Leibler divergence.
The VAE is trained so that $\p{\x@[t]}[\z@[t]]$ and $\q{\z@[t]}$ maximize this variational lower bound.
The first term of Eq. (\ref{eq:vae-elbo}) is analytically tractable and the second term can be approximated with a Monte-Carlo algorithm.
The lower bound can be maximized by using a stochastic gradient descent (SGD)~\cite{kingma2014adam}.


\vspace{-0.5mm}
\section{Statistical Speech Enhancement \\ Based on Combination of VAE and NMF}
\vspace{-0.5mm}
This section describes the proposed probabilistic generative model called VAE-NMF, that combines a VAE-based speech model and a NMF-based noise model.
We formulate the generative process of an observed complex spectrogram $\x* \in \setC^{F\times T}$ by formulating the process of a speech spectrogram $\s* \in \setC^{F\times T}$ and a noise spectrogram $\n* \in \setC^{F\times T}$.
The characteristics of speech and noise signals are represented by their priors based on VAE and NMF, respectively.

\vspace{-0.5mm}
\subsection{VAE-based speech model}
\vspace{-0.5mm}
\begin{figure}
  \centering
  \includegraphics[width=0.95\hsize]{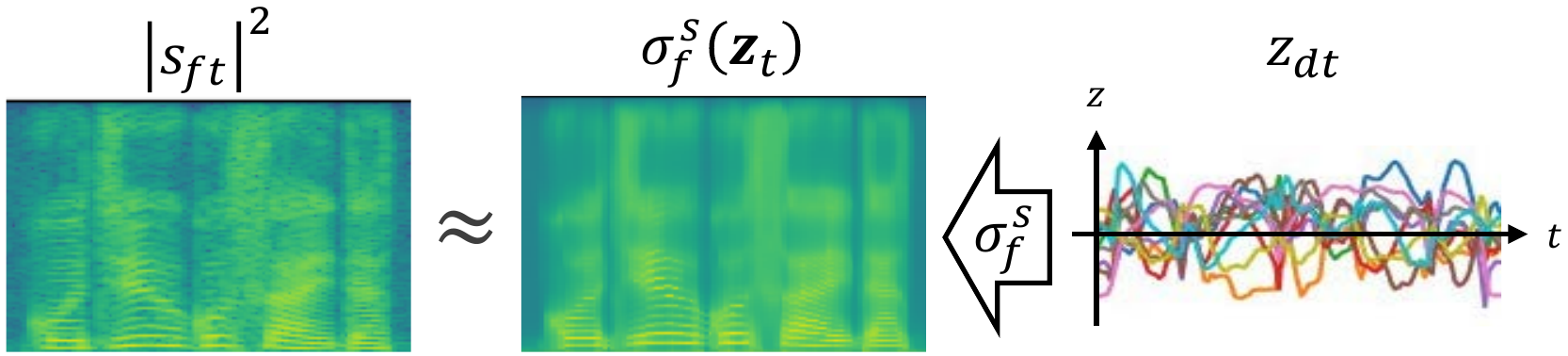}
  \vspace{-4mm}
  \caption{VAE representation of a speech spectrogram.}
  \label{fig:vae}
  \vspace{-3mm}
\end{figure}

In our speech model we assume a frame-wise $D$-dimensional latent variable $\z* \in \setR^{D\times T}$.
Each time-frame of the latent variable $\z@$ is supposed to represent the characteristics of a speech spectrum such as fundamental frequency, spectral envelope, and type of phoneme.
The specific representation of $\z@$ is obtained automatically by conducting the VAE training with a dataset of clean speech spectra.
As in the conventional VAEs,
 we put the standard Gaussian prior on each element of $\z*$:
\begin{align}
  \z \sim \distnormal{0}{1}. \label{eq:p(z)}
\end{align}

Since the speech spectra are primarily characterized by its power spectral density (PSD),
  it follows a zero-mean complex Gaussian distribution whose variance parameter is formulated with $\z*$ (Fig.~\ref{fig:vae}):
\begin{align}
  \s \sim \distcmpnormal{0}{\sigma^s_f(\z@)} \label{eq:p(s|z)},
\end{align}
where $\distcmpnormal{\mu}{\sigma}$ is a complex Gaussian distribution with mean parameter $\mu$ and variance parameter $\sigma$.
$\sigma^s_f(\z@) : \setR^D \rightarrow \setRp$ is a nonlinear function representing the relationship between $\z*$ and the speech signal $\s*$.
This function is formulated by using a DNN and obtained by the VAE training.

\vspace{-0.5mm}
\subsection{Generative model of mixture signals}
\vspace{-0.5mm}
In our Bayesian generative model, the input complex spectrogram $\x* \in \setC^{F\times T}$ is represented as the sum of a speech spectrogram $\s*$ and a noise spectrogram $\n*$:
\begin{align}
  \x = \s + \n.
\end{align}
We put the VAE-based hierarchical prior model (Eqs.~(\ref{eq:p(z)})  and (\ref{eq:p(s|z)})) on the speech spectrogram $\s*$.
On the other hand, we assume that the PSD of the noise spectrogram is low-rank and put an NMF-based prior model on it.
More specifically, the PSD of a noise spectrogram can be represented as the product of $K$ spectral basis vectors $\w* = [ \w@[1], \ldots, \w@[K]] \in \setRp^{F\times K}$ and their activation vectors $\h* \in \setRp^{K\times T}$.
The zero-mean complex Gaussian distribution is put on each TF bin of the noise spectrogram $\n*$ as follows:
\begin{align}
  \n \sim \distcmpnormal{0}{\sum_k \w \h}.
\end{align}
For mathematical convenience, we put conjugate prior distributions on $\w*$ and $\h*$ as follows:
\begin{align}
  \w \sim \distgamma{a_0}{b_0}, \hspace{5mm}\h \sim \distgamma{a_1}{b_1},
\end{align}
where $\distgamma{\alpha}{\beta}$ is a gamma distribution with the shape parameter $\alpha > 0$ and the rate parameter $\beta > 0$; $a_0$, $b_0$, $a_1$, and $b_1$ are hyperparameters that should be set in advance.

By marginalizing out the speech and noise complex spectrograms $\s*$ and $\n*$,
  we obtain the following Gaussian likelihood:
\begin{align}
  \left. \x \right| \w*, \h*, \z* \sim \distcmpnormal{0}{\sum_k \w \h + \sigma^s_f(\z@)}.
\end{align}
Since this likelihood function is independent of the phase term of the input spectrogram $\x*$,
  it is equivalent to the following exponential likelihood:
\begin{align}
  \left. \norm{\x}^2 \right| \w*, \h*, \z* \sim \distexp{\sum_k \w \h + \sigma^s_f(\z@)},  
\end{align}
where $\norm{\x}^2$ is the power spectrogram of $\x*$ and $\distexp{\lambda}$ is the exponential distribution with a mean parameter $\lambda$.
Maximization of the exponential likelihood on a power spectrogram corresponds to minimization of Itakura-Saito divergence,
  which is widely used in audio source separation~\cite{fevotte2009nonnegative, cemgil2009bayesian}.

\vspace{-0.5mm}
\subsection{Pre-training of VAE-based speech model}
\vspace{-0.5mm}
The goal of the pre-training of the VAE-based speech model is
  to find $\p{\s@}[\z@]$
  that maximizes the following marginal likelihood $\p{\s*}$ from the dataset of clean speech signal (denoted by $\s* \in \setC^{F\times T}$ in this subsection):
\begin{align}
  \p{\s*} = \prod_t \int \p{\s@}[\z@] \p{\z@} p\z@.
\end{align}
As stated in Sec.~\ref{sec:vae}, 
  it is difficult to analytically calculate this marginal likelihood.
We approximate it by using the Variational mean-field approximation.
Let $\q{\z*}$ be the variational posterior distribution of $\z*$.
Since $\p{\s*}[\z*]$ is independent from the phase term of the speech spectrogram $\s*$,
  the variational posterior $\q{\z*}$ is defined by ignoring the phase term as follows:
\begin{align}
  \q{\z*} = \prod_{d, t} \q{\z} = \prod_{d, t} \distnormal{\mu^z_d\left(\norm{\s@}^2\right)}{\sigma^z_d\left(\norm{\s@}^2\right)},
\end{align}
where $\norm{\s@}^2$ is the power spectrum of $\s@$ and $\mu^z_d : \setRp^F \rightarrow \setR$ and $\sigma^z_d : \setRp^F \rightarrow \setRp$ are nonlinear functions representing the mean and variance parameters of the Gaussian distribution.
These two functions are defined with DNNs.
The marginal likelihood is approximately calculated as follows:
\begin{align}
  \hspace{-5mm}\log &\p{\s*} \geq -\KL{\q{\z*}}{\p{\z*}} + \E[q]{\log \p{\s*}[\z*]} \\
  &= -\sum_{d, t} \frac{1}{2} \left\{ \left( \mu^z_d(\norm{\s@}^2)\right)^2 + \sigma^z_d(\norm{\s@}^2) -\log \sigma^z_d(\norm{\s@}^2) \right\} \nonumber \\ 
  &\hspace{4mm}+ \sum_{f, t} \E[q]{-\log\sigma^s_f(\z@)-\frac{\norm{\s}^2}{\sigma^s_f(\z@)}} + {\rm const.}
\end{align}
The DNNs for $\sigma^s_f$, $\mu^z_d$, and $\sigma^z_d$ are optimized by using SGD so that this variational lower bound is maximized.

\subsection{Bayesian inference of VAE-NMF}
\setlength\abovedisplayskip{1.5mm}
\setlength\belowdisplayskip{1.5mm}
To enhance the speech signal in a noisy observed signal, we calculate the full posterior distribution of our model: $\p{\w*, \h*, \z*}[\x*]$.
Since the true posterior is analytically intractable, we approximate it with a finite number of random samples by using a Markov chain Monte Carlo (MCMC) algorithm~\cite{bishop2006pattern}.
MCMC alternatively and iteratively samples one of the latent variables ($\w*$, $\h*$, and $\z*$) according to their conditional posterior distributions.

By fixing the speech parameter $\z*$, the conditional posterior distributions $\p{\w*}[\x*, \h*, \z*]$ and $\p{\h*}[\x*, \w*, \z*]$ can be derived with a variational approximation~\cite{cemgil2009bayesian, bishop2006pattern} as follows:
\begin{align}
  &\w|\x*, \h*, \z* \!\sim\! \distgig{a_0}{b_0 \!+\!\! \sum_t \frac{\h}{\auxlmd}}{\sum_t \norm{\x}^2 \frac{\auxphi^2}{\h}}, \\
  &\h|\x*, \w*, \z* \!\sim\! \distgig{a_1}{b_1 \!+\!\! \sum_f \frac{\w}{\auxlmd}}{\sum_f \norm{\x}^2 \frac{\auxphi^2}{\w}}, \\
  &\hspace{-2mm}\auxlmd \!=\! \sum_k \w \h + \sigma^s_f(\z@), \hspace{1mm} \auxphi = \frac{\w \h}{\sum_k \w \h \! +\! \sigma^s_f(\z@)}, 
\end{align}
where $\distgig{\gamma}{\rho}{\tau}\propto x^{\gamma-1} \!\exp(-\rho x - \tau / x)$ is the generalized inverse Gaussian distribution and $\auxlmd$ and $\auxphi$ are auxiliary variables.

The latent variable of speech $\z*$ is updated by using a Metropolis method~\cite{bishop2006pattern}
  because it is hard to analytically derive the conditional posterior $\p{\z*}[\x*, \w*, \h*]$.
The latent variable is sampled at each time frame by using the following Gaussian proposal distribution $\q{\zp@}[\z@]$ whose mean is the previous sample $\z@$:
\begin{align}
  \zp@ &\sim \q{\zp@}[\z@] = \distnormal{\z@}{\sigma \eye},
\end{align}
where $\sigma$ is a variance parameter of the proposal distribution.
This candidate $\zp@$ is randomly accepted with the following probability:
\begin{align}
  a_{\zp@|\z@} &= \min \left( 1, \frac{ \p{\x@}[\w*, \h*, \zp@] \p{\zp@}}
                                      { \p{\x@}[\w*, \h*, \z@] \p{\z@}} \right).
\end{align}

\subsection{Reconstruction of complex speech spectrogram}
In this paper we obtain the enhanced speech with Wiener filtering by maximizing the conditional posterior $\p{\s*}[\x*, \w*, \h*, \z*]$.
Let $\hat{\s*} \in \setC^{F\times T}$ be the speech spectrogram that maximizes the conditional posterior.
It is given by the following equation:
\begin{align}
  \hat{s}_{ft} = \frac{\sigma^s_f(\z@)}{\sum_k \w\h + \sigma^s_f(\z@)} \x. \label{eq:wf}
\end{align}
We simply use the mean values of the sampled latent variables as $\w*$, $\h*$, and $\z*$ in Eq.~(\ref{eq:wf}).

\section{Experimental Evaluation}
\vspace{-0.5mm}
This section reports experimental results with noisy speech signals whose noise signals were captured in actual environments.
\vspace{-1.5mm}
\subsection{Experimental settings}
\vspace{-0.5mm}

To compare VAE-NMF with a DNN-based supervised method,
  we used CHiME-3 dataset~\cite{barker2015third} and DEMAND noise database\footnote{\url{http://parole.loria.fr/DEMAND/}}.
The CHiME-3 dataset was used for both the training and evaluation.
The DEMAND database was used for constructing another evaluation dataset for unseen noise conditions.
The evaluation with the CHiME-3 was conducted by using its development set, which consists of 410 simulated noisy utterances in each of four different noisy environments: on a bus (BUS), in a cafe (CAF), in a pedestrian area (PED) and on a street junction (STR).
The average signal-to-noise ratio (SNR) of the noisy speech signals was 5.8\,dB.
The evaluation with the DEMAND was conducted by using 20 simulated noisy speech signals in each of four different noisy environments: on a subway (SUB), in a cafe (CAF), at a town square (SQU), and in a living room (LIV).
We generated these signals by mixing the clean speech signals of the CHiME-3 development set with the noise signals in the DEMAND database.
The SNR of these noisy speech signals was set to be 5.0\,dB.
The sampling rate of these signals was 16\,kHz.
The enhancement performance was evaluated by using the signal-to-distortion ratio (SDR)~\cite{BSSEVAL}.

\begin{figure}
  \centering
  \subfloat[Encoder: {$\q{\z@}[\s@]$}]{\includegraphics[width=0.49\hsize]{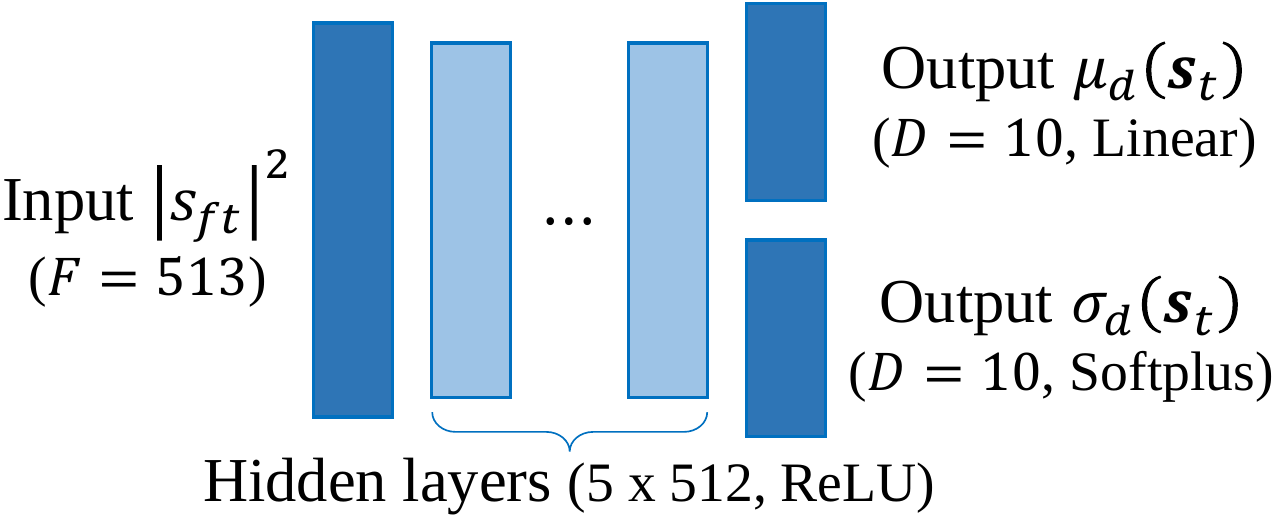}}
  \hfill
  \subfloat[Decoder: {$\p{\s@}[\z@]$}]{\includegraphics[width=0.49\hsize]{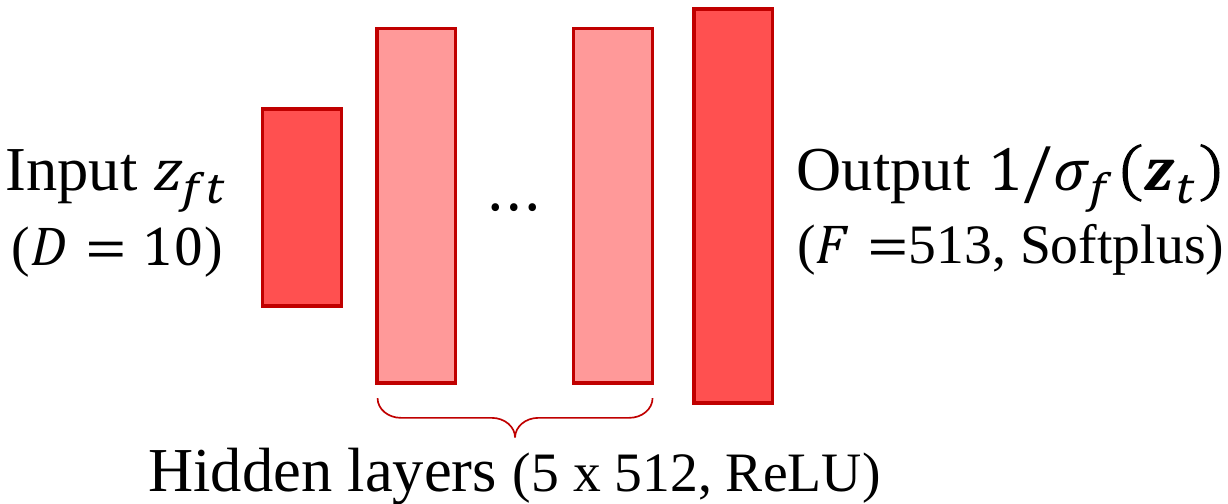}}
  \vspace{-2.5mm}
  \caption{Configuration of the VAE used in the Sec.~4.}
  \label{fig:network}
  \vspace{-3mm}
\end{figure}

To obtain the prior distribution of speech signals $\p{\s@}[\z@]$,
we trained a VAE that had two networks of $\p{\s@}[\z@]$ and $\q{\z@}$ as shown in Fig.~\ref{fig:network}.
The dimension of the latent variables $D$ was set to be 10.
The training data were about 15 hours of clean speech signals in the WSJ-0 corpus~\cite{garofalo2007csr}.
Their spectrograms were obtained with a short-time Fourier transform (STFT) with a window length of 1024 samples and a shifting interval of 256 samples.
To make the prior distribution robust against a scale of the speech power,
we randomly changed the average power of the spectrogram between 0.0 and 10.0 at each parameter update.

The parameters for VAE-NMF were as follows.
The number of bases $K$ was set to be $5$.
The hyperparameters $a_0$, $b_0$, $a_1$, $b_1$, and $\sigma$ were set to be $1.0$, $1.0$, and $1.0$, $K/scale$, and $0.01$, respectively.
The $scale$ represents the empirical average power of the input noisy spectrogram.
After drawing 100 samples for burn-in,
  we drew 50 samples to estimate the latent variables.
These parameters had been determined empirically.
The latent variables of noise $\w*$ and $\h*$ were randomly initialized.
Since the latent variable of speech $\z*$ depends on the initial state,
  the initial sample was drawn from 
  $\q{\z@}[\s@]$ by setting the observation $\x@$ as the speech signal $\s@$.

We compared VAE-NMF with a DNN-based supervised method and the unsupervised RPCA.
We implemented a DNN that outputs IRMs (DNN-IRM).
It had five hidden layers with ReLU activation functions.
It takes as an input 11 frames of noisy 100-channel log-Mel-scale filterbank features and predicts one frame of IRMs%
\footnote{SDRs were evaluated by dropping 2048 samples (5 frames) at both ends.}.
We trained DNN-IRM with the training dataset of CHiME-3, which was generated by using the WSJ-0 speech utterances and noise signals.
The noise signals were recorded in the same environments as those in the evaluated data.

\subsection{Experimental results}
The enhancement performance is shown in Tables~\ref{tab:sdr1} and \ref{tab:sdr2}.
In the experiments using the CHiME-3 test set (Table~\ref{tab:sdr1}),
  DNN-IRM, which was trained using the noisy data recorded in the same environments at the test data,
  yielded the highest average SDR.
The proposed VAE-NMF achieved higher SDRs than RPCA in all conditions and even outperformed the supervised DNN-IRM in BUS condition without any prior training of noise signals.
From the results obtained using the test set constructed with the DEMAND noise data,
  we can see that VAE-NMF outperformed the other methods in all the conditions.
The noise data in DEMAND is unknown to DNN-IRM trained using the CHiME-3 training set,
  and its enhancement performance deteriorated significantly.
These results clearly show the robustness of the proposed VAE-NMF against various types of noise conditions.

The SDR performance of VAE-NMF for the CAF condition in the DEMAND test set
  was lower than those for the other conditions.
In this condition, the background noise contained conversational speech.
Since VAE-NMF estimates speech component independently at each time frame,
  the background conversations were enhanced at the time frames where the power of the target speech was relatively small.
This problem would be solved by making the VAE-based speech model to maintain time dependencies of a speech signal.
The variational recurrent neural network~\cite{chung2015recurrent} would be useful for this extension.

\begin{table}[t]
  \centering
  \caption{Enhancement performance in SDR for CHiME-3 dataset}
  \label{tab:sdr1}
  \footnotesize
  \vspace{0.1mm}
  \setlength\belowrulesep{1.1pt}
  \setlength\aboverulesep{1.1pt}
  \begin{tabular}{l|c|rrrr}
    \toprule
    Method            & Average &  BUS &  CAF &  PED &  STR \\
    \midrule
    VAE-NMF           & 10.10 &  {\bf 9.47} & 10.62 & 10.93 &  9.39 \\
    DNN-IRM           & {\bf 10.93} &  8.92 & {\bf 11.92} & {\bf 12.92} &  {\bf 9.95} \\
    RPCA              &  7.53 &  6.13 &  8.10 &  9.13 &  6.77 \\
    Input             &  6.02 &  3.26 &  7.21 &  8.83 &  4.78 \\
    \bottomrule
  \end{tabular}
  \vspace{-2mm}

  \caption{Enhancement performance in SDR for DEMAND dataset}
  \label{tab:sdr2}
  \vspace{0.1mm}
  \footnotesize
  \begin{tabular}{l|c|rrrr}
    \toprule
    Method            & Average &  SUB &  CAF &  SQU &  LIV \\
    \midrule
    VAE-NMF           & {\bf 11.17} & {\bf 10.56} &  {\bf 9.57} & {\bf 12.38} & {\bf 12.16} \\
    DNN-IRM           &  9.85 &  9.13 &  9.15 & 10.69 & 10.42 \\
    RPCA              &  7.03 &  6.48 &  6.37 &  6.99 &  8.28 \\
    Input             &  5.21 &  5.25 &  5.24 &  5.19 &  5.16 \\  
    \bottomrule
  \end{tabular}
  \vspace{-2mm}
\end{table}


\vspace{-2mm}
\section{Conclusion}
\vspace{-2mm}
We presented a semi-supervised speech enhancement method, called VAE-NMF,
  that involves a probabilistic generative model of speech based on a VAE
  and that of noise based on NMF.
Only the speech model is trained in advance by using a sufficient amount of clean speech.
Using the speech model as a prior distribution,
 we can obtain posterior estimates of clean speech by using an MCMC sampler
 while adapting the noise model to noisy environments.
We experimentally confirmed that VAE-NMF outperformed 
 the conventional supervised DNN-based method in unseen noisy environments.

One interesting future direction is to extend VAE-NMF to the multichannel scenario.
Since complicated speech signals and a spatial mixing process can be represented
 by a VAE and a well-studied phase-aware linear model (e.g., \cite{nugraha2016multichannel, heymann2016neural, ozerov2010multichannel}), respectively,
 it would be effective to integrate these models in a unified probabilistic framework.
We also investigate GAN-based training of the speech model
 to accurately learn a probability distribution of speech.

\end{document}